\documentclass{iopjournal}

\begin{document}

\articletype{Paper} %	 e.g. Paper, Letter, Topical Review...

\title{HPGe-Compton Net: A Physics-Guided CNN for Fast Gamma Spectra Analysis via Compton Region Learning}

\author{Yanfeng~Xie$^{1,2,*}$\orcid{0009-0005-3541-9500}, Yiming~Weng$^{1,3}$\orcid{0009-0009-9803-579X} and Soo~Hyun~Byun$^{1,2}$\orcid{0000-0002-0029-5094}}

\affil{$^1$Radiation Sciences Graduate Program, McMaster University, Hamilton, ON, L8S 4K1, Canada}

\affil{$^2$Department of Physics and Astronomy, McMaster University, Hamilton, ON, L8S 4K1, Canada}

\affil{$^3$Health Physics Department, McMaster University, Hamilton, ON, L8S 4L8, Canada}

\affil{$^*$Author to whom any correspondence should be addressed.}
\email{xie22@mcmaster.ca}


\begin{abstract}
High-Purity Germanium (HPGe) detectors have been golden standard for gamma spectrometry in Low-level radioactive waste (LLW) analysis; however, their notable shortcoming is prolonged measurement durations for weak radioactive waste materials. The present study aimed to develop the HPGe-Compton Net, a 1D Physics-Guided CNN to accelerate LLW analysis by taking advantage of the entire response function of a HPGe detector for each radionuclide of interest, in contrast to the traditional methods that analyze only peak regions of the response. This acceleration is supported by two core innovative strategies: (a) Channel-Prompt method, a feature enhancement incorporating additional physical information to guide the model to locate the designated radionuclide; (b) the specially designed database to achieve effective targeted feature learning. The performance evaluation carried out for test data set showed a five times reduction in measurement time compared to a conventional spectral analysis method while maintaining comparable precision. Compton perturbation tests confirmed the model’s “smart” adaptive utilization of the Compton regions. The generalization testing of four LLW samples as the external validation set proved its superior performance in low-count data with an average accuracy of 90\% over 83\% of the traditional method. Future work will focus on upgrading the HPGe-Compton Net for practical applications.
\end{abstract}

\section{Introduction}
Low-level radioactive waste (LLW), predominantly generated by nuclear reactors, hospitals, and scientific research centers, contains modest concentrations of long-lived radionuclides that necessitate isolation and containment to mitigate radiation hazard. Gamma spectrometry enables precise identification of radionuclides and their activities, supporting regulatory classification systems that determine containment durations ranging from hundreds of days to several centuries\cite{CNSC}. Rapid and reliable gamma spectrometric characterization, therefore, is of great importance in optimizing LLW management protocols.

High-Purity Germanium (HPGe) detectors have been employed as the golden standard detection systems for gamma spectrometric measurements in both fundamental and applied experimental nuclear physics thanks to their excellent energy resolution. The detector interactions with incident gamma radiation include photoelectric absorption, Compton scattering, and pair production. The response of a detector, i.e. the distribution of the deposited energy, for a given gamma energy has a unique feature that consists of a full-energy peak from full-energy deposition events, a Compton continuum from partial energy deposition events by Compton scatterings, single and double escape peaks that are produced from the pair production interaction. A typical spectral response is illustrated through experimental measurements using a Cs-137 calibration source (Fig. \ref{fig:1}). Despite the energy of the gamma photon emitted from the source being mono-energetic, the detector records counts across multiple spectral regions:
\begin{figure}[!t]
  \centering
  \includegraphics[width=0.95\columnwidth]{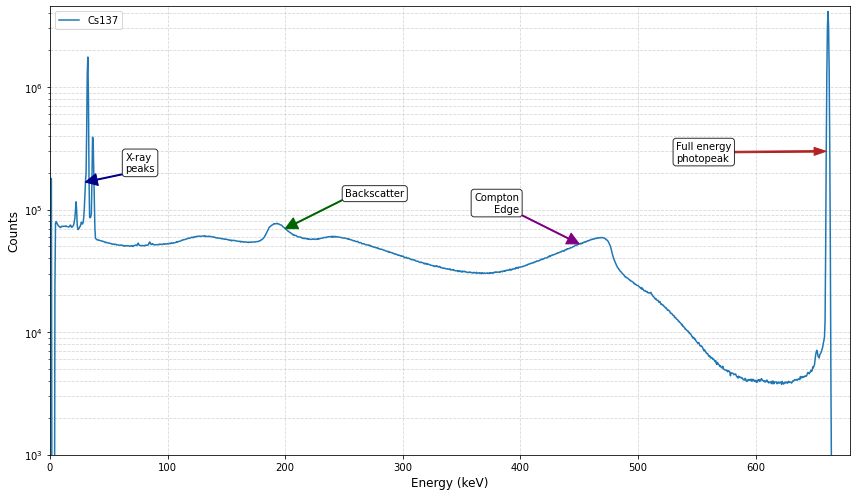}
  \caption{A gamma spectrum collected for measuring Cs-137 mono-energy source.}
  \label{fig:1}
\end{figure}
(a) X-ray peaks: these low-energy bumps are the full-energy peaks for the X-rays emitted by Ba (Cs-137 decay daughter) or characteristic X-rays that were emitted following the photoelectric absorption of the Cs-137 gamma photons by the detector's surroundings or shielding. (b) Compton continuum originates from a Compton scattering, which leads to a partial energy deposition event that is dependent on the scattering angle. This results in an energy distribution with two notable features: the backscatter peak and the Compton edge. If Compton scatterings occur by the detector housing material or shielding at large scattering angles close to 180 degrees, these photons will be scattered back into the detector, producing the backscatter peak via the photoelectric absorption.  The energies of the backscattered gamma photons are less sensitive to the incident gamma energies and are around 200 keV. The Compton edge represents the maximum energy that an incident gamma photon can transfer to a bound electron in the detector in a single scattering event, as described by the Compton scattering equation. (c) Full energy peak, an incident gamma photon has deposited its entire energy in the detector. The full energy peak has a sharp Gaussian-like shape and serves as a nuclide-specific signature, enabling direct radionuclide identification. 

Additional spectral features arise from instrumental and environmental factors, including statistical fluctuation in the number of detected counts, background radiation contributions such as cosmic rays, natural radioactive elements present in building materials, and response modification by the coincidence summing effect, which happens when two or more gamma photons of different energies emitted simultaneously from a cascade gamma decay event from the source make interactions with the detector and produce a single detection pulse that corresponds to the sum of the energies deposited by the cascade gamma photons. 

For decades, gamma spectra analyses have exclusively utilized the full energy peak for HPGe detectors. In operational LLW measurements, the collected spectra usually comprise multiple radionuclides’ responses to the detector with the full energy peaks serving as the most identifiable feature. The activity of each radionuclide is quantitatively analyzed using the full energy peak counts determined from the collected gamma spectrum and the energy dependent peak efficiency of the detector, which was calibrated with experimental measurements or Monte Carlo simulations. By improving the full energy peak quantification method and counting statistics, the final radioactivity evaluation can be improved synchronously.

Accurate full energy peak quantification necessitates advanced curve-fitting algorithms. Popular curve fitting methods employ a modified Gaussian function with about four free shape parameters to characterize its width variation, connected tail-like components, background subtraction, etc. Shape parameters are determined via nonlinear least-squares fit requiring multiple calibration measurements \cite{Canberra}. As gamma detection counts follow Poisson statistics, fitting precision is fundamentally governed by counting statistics, with count uncertainty defined as the square root of total counts: higher total counts yield lower relative uncertainty and enhanced measurement precision. In low-activity LLW samples (\textless10 Bq/kg), measurement durations often extend to several days to achieve good counting statistics for each sample, leading to a significant increase in the total measurement duration. However, the fundamental physics of gamma-ray interactions suggests that underutilized spectral regions, such as the Compton continuum, contain complementary analytical information. Comprehensive utilization of these features could theoretically reduce the required measurement time. In 1961, Salmon \cite{LLS} developed the Library Least Squares (LLS) method to analysis the entire gamma spectra rather than the full energy peak only. LLS has been shown to effectively reduce the uncertainty in neutron capture prompt gamma analysis \cite{PGNAA}. However, LLS is more suitable for low energy resolution gamma detector like NaI detectors, due to its complicated logic \cite{Uncertainty}. Recent advancements in sparse spectral unmixing methodologies also aim to enhance quantitative accuracy by analyzing full spectral datasets regions \cite{spectral unmixing}, \cite{spectral unmixing2}. Despite this potential, general industrial applications remain limited by operational complexity and marginal time-reduction benefits. 

Deep learning models have been extensively investigated for HPGe spectral analysis to simplify specialized analytical workflows \cite{uranium concentration}, \cite{isotopic determination}, though existing implementations have yet to surpass traditional curve-fitting methodologies in analytical precision. In these studies, there are complaints about insufficient datasets. To address these issues, Physics-Guided neural networks (PGNN) can be used to reduce the demand for dataset size by fully utilizing the known physical knowledge of gamma spectra. The physics knowledge in this work specifically represents the gamma decay energy of radionuclides and the usefulness of feature regions like the Compton Edge. Physics knowledge can be fed into machine learning by enhancing features based on additional physics knowledge as input to the model or a specifically defined loss function based on physics differential equations, expecting the model to achieve more effective learning with the same database \cite{PGNN review}. PGNN has had a significant impact in various industrial research fields \cite{PGNN Power}, \cite{PGNN Lake}, and has demonstrated superiority in situations where data is insufficient, relying on physical knowledge rather than solely relying on statistical patterns induced from datasets. 

In collaboration with Laurentis Energy Partners (LEP) and McMaster University Health Physics, our research group has developed HPGe-Compton Net—an 1D Physics-Guided CNN network architecture trained to extract accurate information from both Compton regions and the full energy peak, by incorporating a Channel-prompt part as additional physics information. Methodology section first introduces basic instrument setup and difficulties necessitating the exploration of innovative methods, such as the Channel-prompt method, and then describes how the specially designed database with a clean Compton edge region and HPGe-Compton Net structure is built. Result \& Discussion section not only strictly evaluates the model performance against traditional curve fitting methods, but also persuasively verifies the physics-guiding abilities provided by the Channel-prompt method. Even more notable is that the effectiveness of the specially designed database, guiding the deep learning model, achieves Compton-edge-targeted feature learning, and it enhances the interpretability of the model.

\section{Methodology}
\subsection{Basic Setup}
The HPGe gamma spectrometer consists of an ultra low-background lead shield (model 777A, Mirion) \cite{Mirion 777}, a broad-energy Germanium detector (model BEGe, Mirion), covering the energy region of 3 keV to 3 MeV \cite{Mirion BEGe}, and a digital pulse processor (model Lynx DSA, Mirion). The digital pulse processor has 16k channel histogramming memory as a multichannel analyzer (MCA), and 4096 channels were used for gamma spectral data collection in the present study. The spectrometer setup, its energy and efficiency calibrations were carried out using I-129, Co-60, and Cs-137 calibration sources. The spectrometer has shown consistent and reliable performance for radioactivity monitoring for samples taken from the McMaster Nuclear Reactor (MNR) and radioactive waste classification over multiple years of operation \cite{LionelManual}. The spectrum analysis software (Genie2000, Mirion) was employed as the benchmark spectrum analysis throughout this study \cite{Canberra}. To quote the accuracy of the deep learning method against Genie2000, this work employs the percentage accuracy value: one minus the percentage residuals. In this study, the use of average percentage accuracy across a large number of similar measurements effectively reflects the overall performance of the model, without the need to estimate uncertainty for each individual point. Since uncertainty in deep learning predictions cannot be derived from first principles, the performance evaluation (e.g., average accuracy) inherently captures the statistical fluctuation. As the present study does not aim to address radionuclide presence/absence determination, scenarios involving zero counts from Genie2000 in accuracy calculations are excluded from consideration. Genie2000’s full energy peak analysis accuracy is governed by counting statistics: $\sim$98\% accuracy requires $\geq2,500$ net counts in the full energy peak region; $\sim$99\% accuracy necessitates $\geq10,000$ net counts. The primary objective of the present study centers on deploying deep learning techniques to analyze spectra acquired under reduced measurement durations while maintaining accuracy levels equivalent to those achieved through conventional full-duration measurements. 

The deep learning training database was constructed as follow: For each sample, spectral data acquired under reduced measurement durations serve as model inputs, while the corresponding full energy peak information analyzed by Genie2000 from full-duration measurements—divided by an acceleration factor—provide the target outputs. The acceleration factor is defined as the ratio of full measurement duration to reduced measurement duration, representing the expected deep learning analysis acceleration performance. This acceleration factor is constrained between 3 and 10 in our experiments. Measurement durations per sample range from 30 minutes to multiple days, ensuring full energy peak net counts exceed 20,000 for good statistical precision.

Operational constraints, including health physics regulations and detector limitation, have limited data acquisition to 150 spectra from 45 samples over nine months. The recorded count rate varies from 3 to around 1400 counts per second, and the measurement duration ranges from 3 min to longer than 4 days. These samples exhibit significant diversity, including radioactive water samples from MNR, contaminated lab coats, contaminated shoes, tooth samples from the cancer center, dirt cleanup collections from a hot cell entrance, and other unknown radioactive materials in plastic packaging etc. Each HPGe spectrum includes 4096 channels, and there are multiple peaks in most spectra. Two primary technical challenges are encountered for ML spectral analysis: (a) Data scarcity: Training 1D convolutional networks on 4096-length spectral vectors with \textless200 samples are inadequate in principle; (b) Localization of region of interests (ROI): Developing reliable methods to identify full energy peak/Compton regions for AI’s attention remains unresolved. Preliminary hybrid networks supporting both spectral data and ROI annotations as input led to a very low accuracy of 30\%.

\subsection{Channel-Prompt Section}
To address these challenges, we implemented a creative feature augmentation approach by appending a Channel-Prompt section to the original 4096-channel spectrum, guiding CNN to put its attention on ROI, with known physics information: the incident gamma energy (e.g., Cs-137: 662 keV, Co-60: 1173 \& 1332 keV). This method involves concatenating the spectrum data with 128 repetitions of the target nuclides' full energy peak centroid positions, resulting in a 4224-channel input vector. By generating Channel-Prompt sections for multiple target full energy peak energies, we expanded the original 150 spectra into 3,000 training datasets. A 1D convolutional neural network (CNN) trained on this augmented data achieved 70\% accuracy in test dataset evaluations, validating the Channel-Prompt concept despite underperforming Genie2000’s benchmark. 

Two critical limitations explain this performance gap: (a) Data insufficiency: The 3,000-sample dataset remains inadequate for robust training of large-size (4224-channel) spectral data; (b) Compton contamination: In multi-nuclide spectra, Compton regions of target radionuclides are often obscured by overlapping full energy peaks from other radionuclides. With counts per channel in Compton region typically 10–1,000× lower than those in the corresponding full energy peak region, Poisson noise from the adjacent full energy peak frequently exceeds Compton signal magnitudes, complicating information obtaining. To enhance model accuracy, new datasets are required to be specifically designed to preserve uncontaminated Compton regions through controlled sample compositions.

\subsection{Database Build}
Six candidate nuclides are selected based on their prevalence in LLW: Cs-137 (662 keV), Co-60 (1173 \& 1332 keV), Na-22 (1275 keV), Am-241 (59 keV), Eu-152, and Eu-154. Practical considerations refined this selection: (a) Eu-152/154 were excluded due to their multiple full energy peaks (\textgreater4 per radionuclide), which enables robust activity determination through peak weighting, reducing the need for accelerated measurement. (b) Am-241 was omitted because its low-energy gamma photons(26 \& 59 keV) produce negligible Compton regions within the 0–2048 keV detection range. (c) Co-60 was retained despite spectral complexity (1173 keV full energy peak overlapping 1332 keV Compton edge) due to its great LLW prevalence and predictable 1173 \& 1332 keV gamma intensity ratio, theoretically resolvable with adequate training data.
\begin{figure}[ht]
  \centering
  \includegraphics[width=0.95\columnwidth]{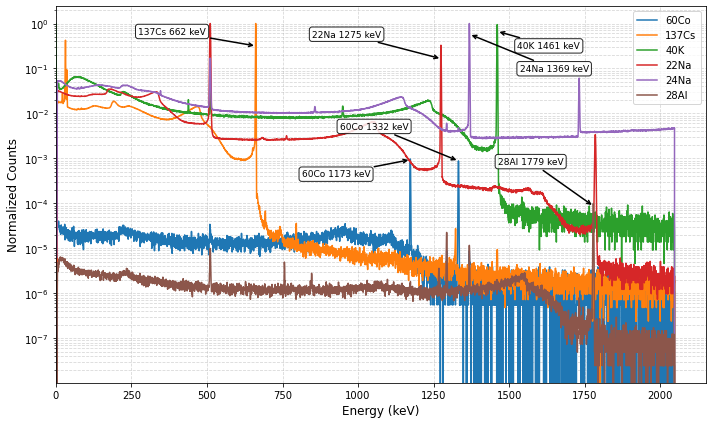}
  \caption{Gamma spectra collected for measuring Cs-137, Co-60, K-40, Na-22, Na-24, and Al-28 sources.}
  \label{fig:2}
\end{figure}
The finalized calibration suite comprises Cs-137, Co-60, Na-22, and three supplementary radionuclides: Na-24, Al-28, and K-40. Na-24 (1369 \& 2754 keV) and Al-28 (1779 keV) are produced via neutron activation in the McMaster Nuclear Reactor rabbit system, though they are not present in real LLW analysis due to their short half-life, their spectra are good material for CNN to learn the full energy peak and Compton region features among various energy \cite{reactor}, K-40 (1461 keV), a ubiquitous background source requiring precise differentiation between sample-contained and environmental contributions. Natural potassium chloride (500 g) provides measurable K-40 activity due to its 0.012\% natural isotopic abundance.

Figure \ref{fig:2} displays normalized spectra of the six target nuclides (Co-60, Cs-137, K-40, Na-22, Na-24, Al-28), with distinct colors representing individual radionuclides. Vertical “line-like” sharp full energy peak demonstrates the HPGe detector’s superior energy resolution.

The spectra of these six nuclides were acquired with sufficient measurement duration to ensure \textgreater1,000 counts per channel in the Compton regions, achieving full energy peak area uncertainties below 0.5\%. Linear scaling adjusted peak areas to the 50$\sim$5,000 count range—the operational window where deep learning acceleration might be feasible. Poisson noise was added to the counts of each channel to make the data more realistic. To enhance dataset realism, low-energy spectral artifacts (50$\sim$700 keV) with randomized intensities were superimposed on each spectrum. These artifacts were derived from real LLW measurements and low energy calibration sources like Cd-109. Contamination thresholds were set to avoid critical analytical regions: (a) Cs-137 datasets: Contaminants restricted to \textless300 keV to preserve Compton edge integrity; (b)	Other nuclides: Contaminants (including Cs-137) limited to \textless700 keV.

All measurements maintained \textless2\% dead time to ensure valid spectral superposition calculation. Through the Channel-Prompt method, we generated 875,000 datasets exhibiting both scale and combinatorial diversity. 20\% of the datasets are left as test datasets.

\subsection{HPGe-Compton Net Structure}
HPGe-Compton Net process 1D sequential data with mixed physics information, including spectra and the Channel-prompt section, aimed at comprehensive analysis the entire spectra, especially the designated full energy peak and Compton region to output the accurate full energy peak area counts. 

\begin{figure*}[ht]
  \centering
  \includegraphics[width=\textwidth]{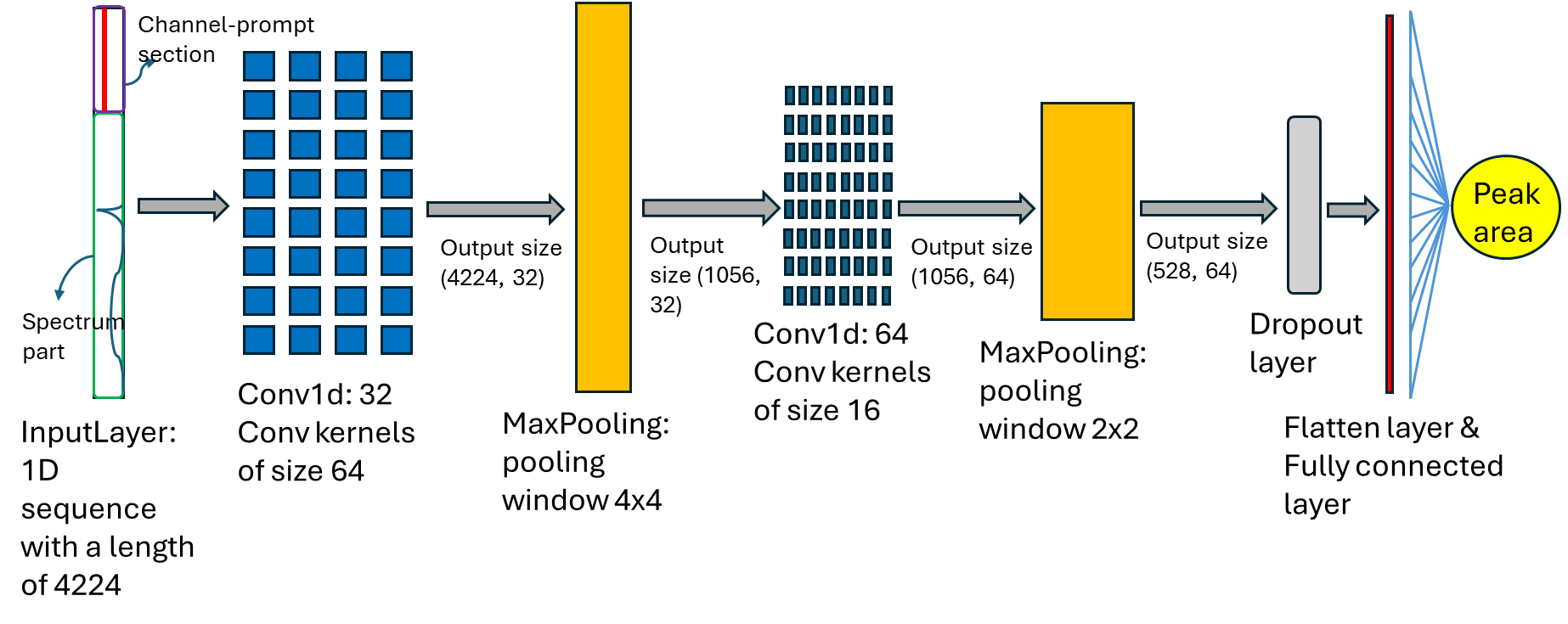}
  \caption{HPGe-Compton Net Structure illustration.}
  \label{CNN}
\end{figure*}

Due to the availability of sufficient and high-quality datasets this time, as well as the previous 3000 dataset used for pre-training, the network structure obtained from this fine-tuning is not too complex. Figure \ref{CNN} presents the HPGe-Compton Net structure developed on TensorFlow\cite{TensorFlow}: its framework belongs to a typical one-dimensional CNN: The input layer accepts a one-dimensional vector and is followed by the first convolutional layer utilizing 32 filters with a kernel size of 64, employing the ReLU activation function, designed for coarse feature extraction. This is followed by a max-pooling layer with a pool size of 4 to improve computational efficiency. The second convolutional layer increases the number of filters to 64, with a reduced kernel size of 16, also employing the ReLU activation function, for fine spectral feature detection. This is then followed by a max-pooling operation with a pool size of 2. To mitigate overfitting and enhance the generalization ability of the model, a dropout layer with a dropout rate of 0.3 is incorporated after the second pooling layer. Following the convolutional layers, the extracted feature maps are flattened into a one-dimensional vector and passed through a fully connected layer consisting of 128 neurons. Finally, the output layer consists of a single neuron to predict a continuous numerical value, the full energy peak counts. In the present study, the mean absolute error (MAE) is adopted as the chosen loss function.

\section{Results \& Discussion}
\begin{figure}[!t]
  \centering
  \includegraphics[width=0.95\columnwidth]{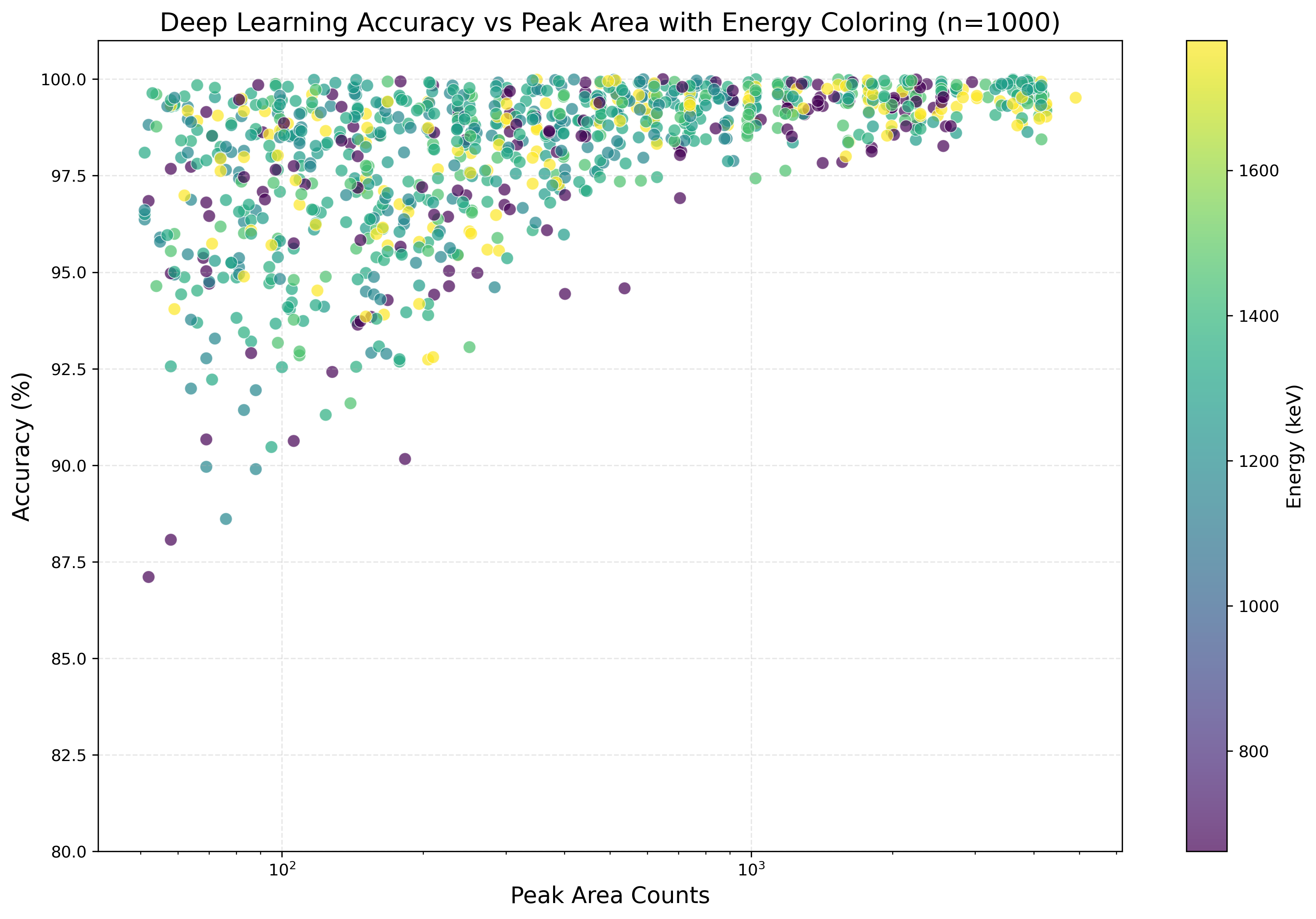}
  \caption{The relationship between HPGe-Compton Net prediction accuracy, full energy peak area counts, and the gamma energy.}
  \label{f4}
\end{figure}

The performance of HPGe-Compton Net was evaluated using 1,000 randomly selected data from test dataset. The model achieved an average accuracy of 98.1\% across full energy peak areas spanning 50$\sim$5,000 counts, demonstrating robust analytical capability within this operational range. Fig. \ref{f4} illustrates the relationship between prediction accuracy and full energy peak counts for these 1,000 test cases, with gamma energy represented by a color gradient. Accuracy improvements correlate strongly with the increase of full energy peak counts, exhibiting tighter clustering around higher precision values at elevated count levels, which is consistent with statistical principles. No discernible energy dependence was observed, confirming the model’s applicability across broad gamma energy ranges.

\begin{figure*}[ht]
  \centering
  \includegraphics[width=\textwidth]{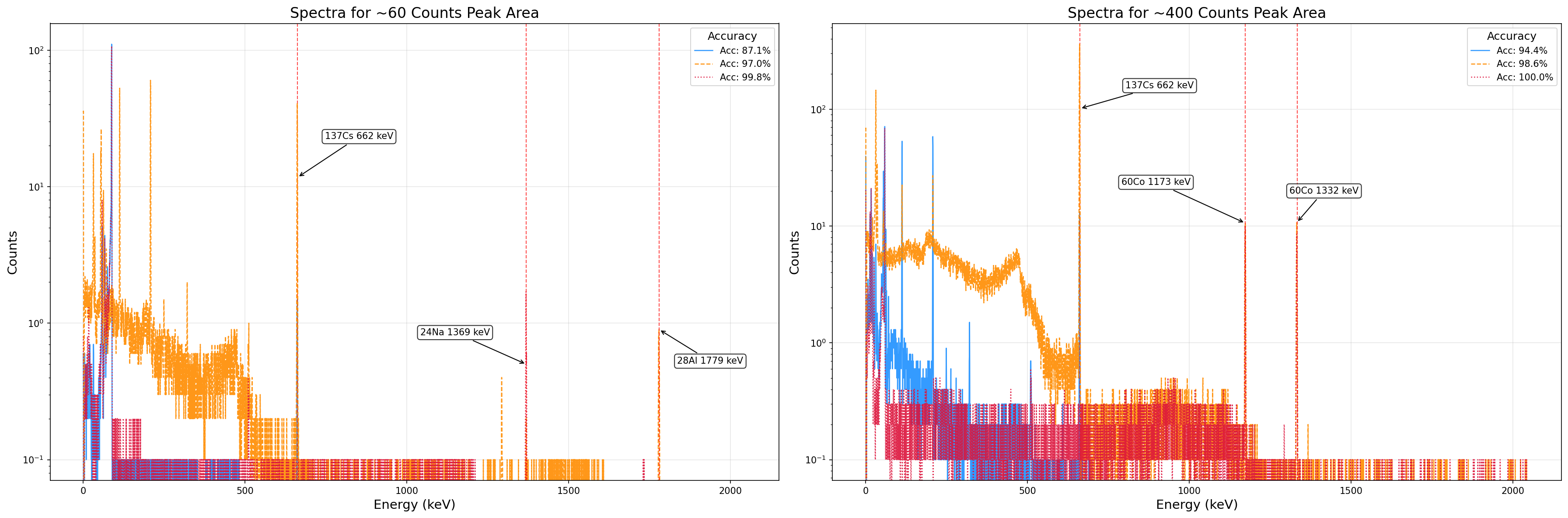}
  \caption{Three spectra visualization (high, medium, low accuracy) for around 60 counts peak area; Another three spectra visualization (high, medium, low accuracy) for around 400 counts peak area.}
  \label{f5}
\end{figure*}

In order to further compare the analytical ability of HPGe-Compton Net for spectra with different peak areas, in Fig.\ref{f5}, six sets of data were selected from 1000 test cases to visualize the input spectra and model performance in the case of high ($\sim$400) and low ($\sim$60) peak areas. The vertical red dashed line is used to indicate the designated full energy peak position. The input spectra have diverse shapes and complex compositions in the low-energy region, which is consistent with the dataset building. As the peak area increases, accuracy generally improves.

\begin{figure}[!t]
  \centering
  \includegraphics[width=0.95\columnwidth]{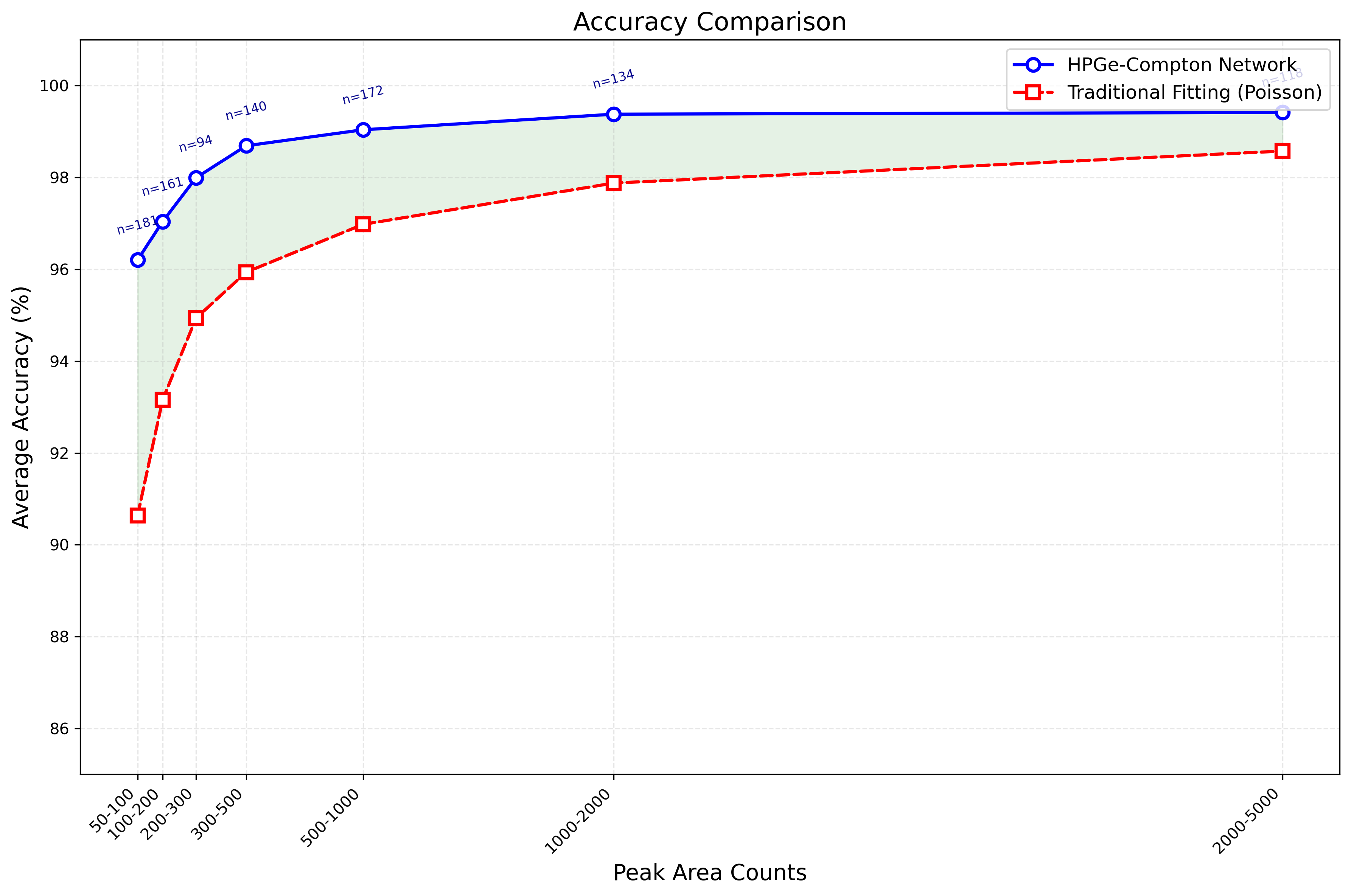}
  \caption{Average accuracy comparison between HPGe-Compton Net and Traditional curve fitting.}
  \label{f6}
\end{figure}

For direct comparison with traditional curve fitting, the test data were divided into seven groups based on full energy peak count ranges (Fig. \ref{f6}). Group sizes are annotated above each data cluster. The blue curve denotes the group-averaged accuracy of HPGe-Compton Net, while the red dashed curve represents the averaged accuracy of the Genie2000 fitting, which is governed by Poisson statistics. Light gray auxiliary grid lines in Fig. \ref{f6} indicate count thresholds for matching accuracy levels: if the required accuracy is 96\%, traditional curve fitting requires 300$\sim$500 peak counts while HPGe-Compton Net only requires 50$\sim$100 peak counts, accelerating the measurement fivefold. If the required accuracy is 98\%, around fivefold acceleration of the measurement can also be provided by HPGe-Compton Net: Traditional methods require 1000$\sim$2000 counts while 200$\sim$300 counts are sufficient for HPGe-Compton Net. In general, HPGe-Compton Net can greatly shorten the required measurement duration while keeping the required accuracy and provide the technical potential to significantly improve the efficiency of LLW analysis.

For industrial deployments of HPGe-Compton Net, a comprehensive performance evaluation must be carried out to validate its performance and meet the acceptance criteria required in LLW analysis. While the database design emphasizes Compton region utilization as the hypothesized acceleration mechanism, critical questions remain unresolved: Does the model really leverage Compton region data, or does it employ enhanced peak-fitting strategies? Could performance gains stem from overfitting rather than valid physics-based learning?

\begin{figure*}[ht]
  \centering
  \includegraphics[width=\textwidth]{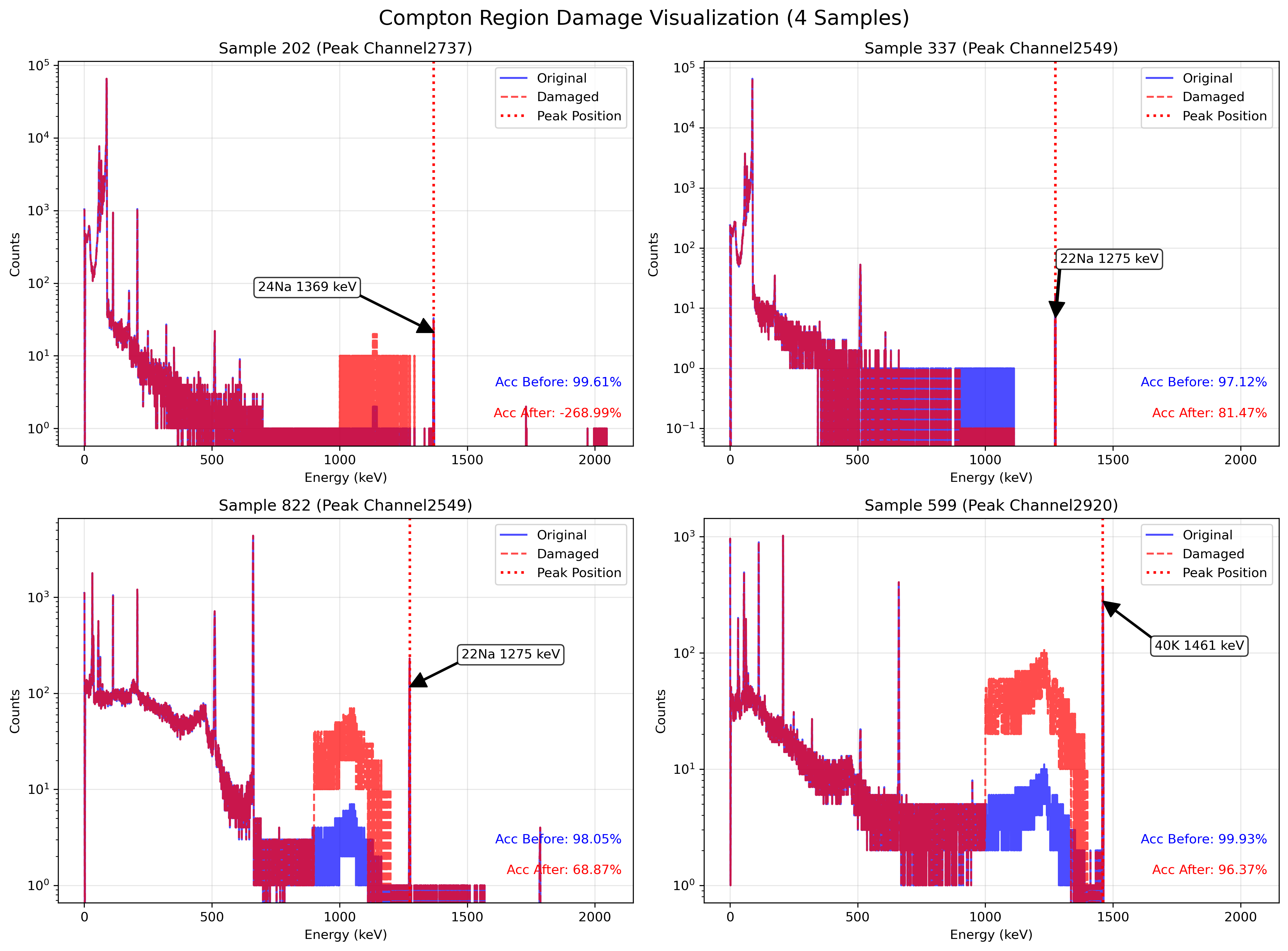}
  \caption{Compton region damage and spectra visualization of Sample 202, Sample 337, Sample 822, and Sample 599}
  \label{f7}
\end{figure*}

A test for Compton region damage was devised to investigate model dependence on Compton region features. Artificial noise was introduced to critical Compton edge regions by randomly scaling channel counts to 10\% or 1000\% of the original values. Four representative samples (\#202, \#337, \#822, \#599) from the 1,000-test dataset were analyzed in Fig. \ref{f7}: (a) Undamaged spectra: Blue curves with full energy peak centroids marked by vertical dotted lines; (b) Damaged spectra: Red dashed curves with distorted Compton regions. Notable noise artifacts (red fluctuations) appear in high-count samples (\#822, \#599), while low-count samples (\#202, \#337) show non-visible distortion. Damaged spectra analysis accuracy dropped significantly for all samples except \#599 (high full energy peak counts), suggesting the model prioritizes full energy peak data when count statistics permit.

\begin{figure}[!t]
  \centering
  \includegraphics[width=0.95\columnwidth]{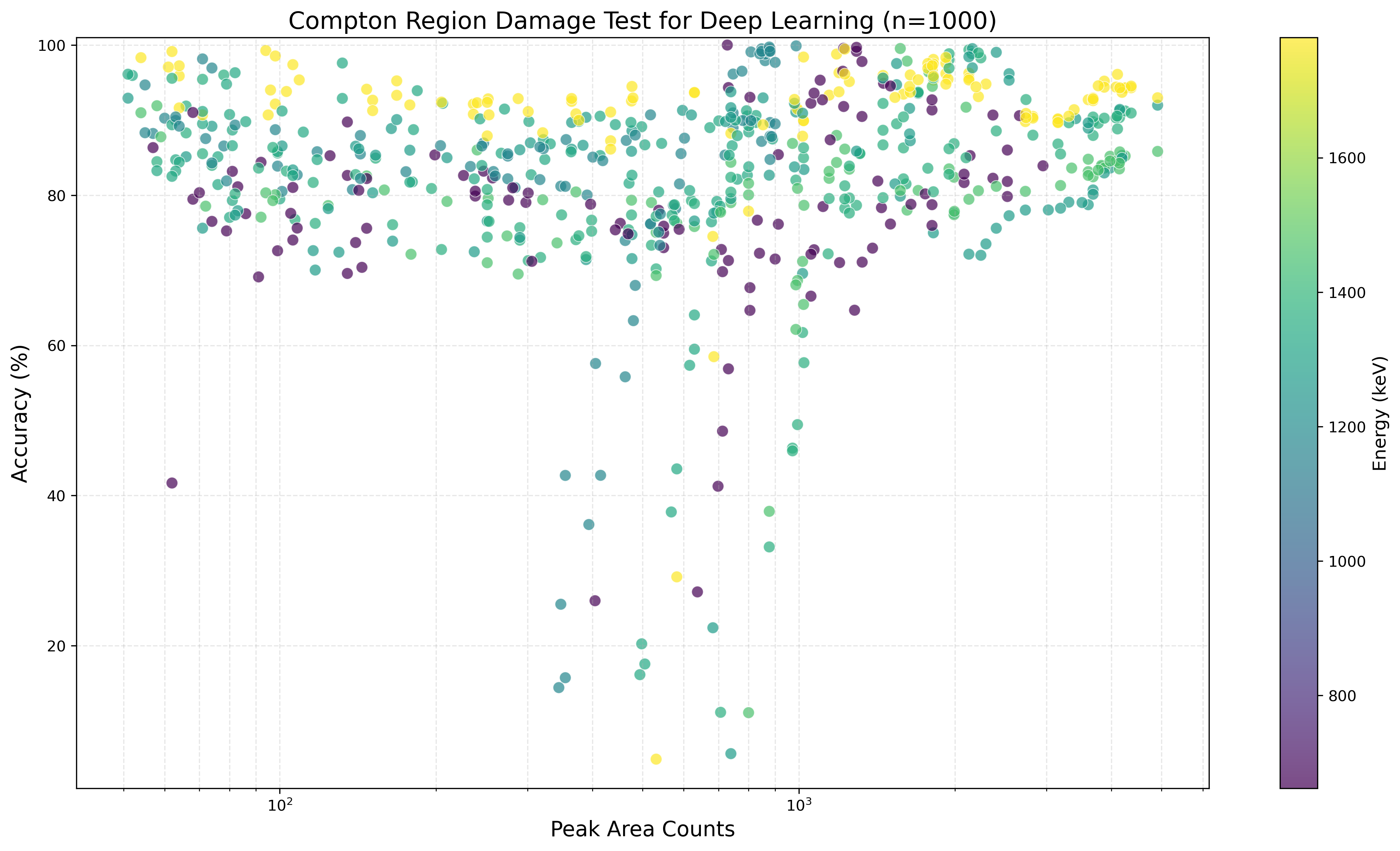}
  \caption{The relationship between HPGe-Compton Net prediction accuracy after applying Compton region damage, full energy peak area counts, and the gamma energy.}
  \label{f8}
\end{figure}

\begin{figure*}[ht]
  \centering
  \includegraphics[width=\textwidth]{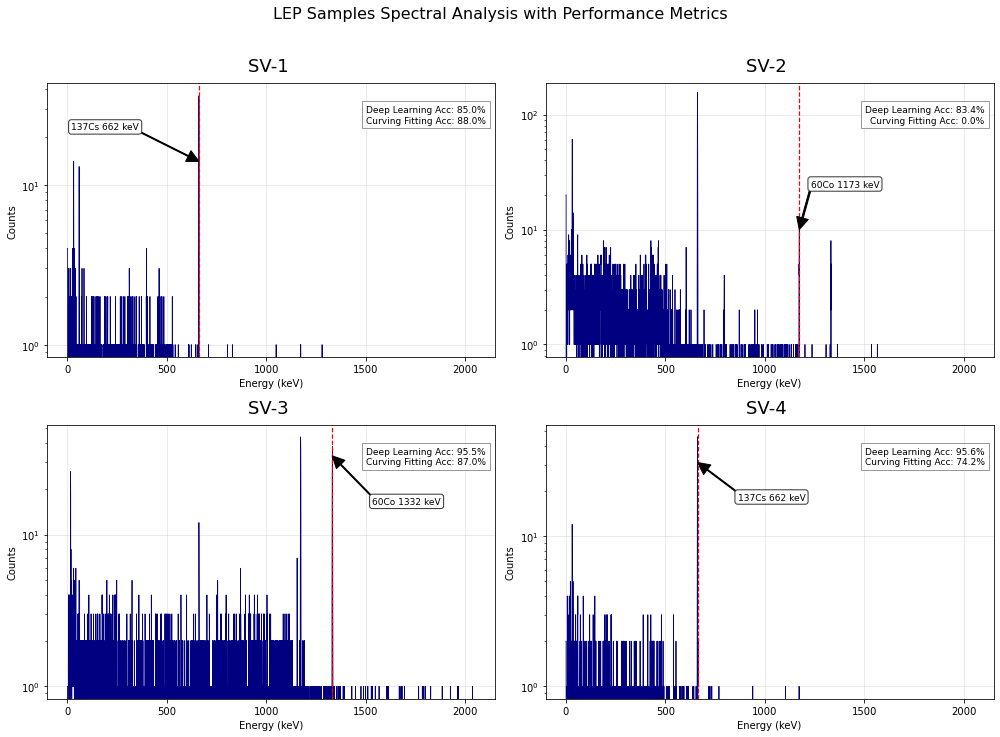}
  \caption{Comparison between HPGe-Compton Net and Genie2000 curve fitting method on LEP SV-series samples gamma spectral analysis.}
  \label{f9}
\end{figure*}

To further examine HPGe-Compton Net’s analytical prioritization, the same 1,000 test spectra from Fig. \ref{f4} were reanalyzed after applying Compton region damage (Fig. \ref{f8}). Compared to the original distribution, Fig. \ref{f8} exhibits a distinct "T"-shaped pattern: significant accuracy degradation occurs exclusively in samples with full energy peak counts between 300$\sim$1,000. This indicates HPGe-Compton Net dynamically adjusts its weighting of full energy peak and Compton region features based on spectral statistics. Two possible operational regimes can explain this behavior: (a) Low-count spectra (\textless300 full energy peak counts): Insufficient Compton counts preclude meaningful learning, forcing reliance on full energy peak data; (b) High-count spectra (\textgreater1,000 counts): Full energy peak statistics alone suffice for precise analysis, reducing Compton region dependence. The "T"-shaped degradation pattern conclusively demonstrates the Physics-Guided CNN model’s intelligent utilization of Compton region information—leveraging it only when statistically advantageous. This Compton region damage test proves that controlled uncontaminated-Compton-edge-region database effectively guides CNN to learn the designated nuclide’s gamma full energy peak and the targeted Compton features.

As sated in the database building chapter, HPGe-Compton Net imposes three critical requirements for input spectra: (a) The target nuclide must be among Co-60, Cs-137, K-40, Na-22, Na-24, or Al-28; (b) The corresponding Compton edge region must remain uncontaminated by extraneous full energy peak; (c) No other nuclides exist in the higher energy region. As of the completion of HPGe-Compton Net training, there are 10 LLW samples from LEP available in our measurement facility, and 4 of them (labeled SV1$\sim$4) basically meet the three requirements above. These samples contained radioactive nuclides not included in the database, for example, like Co-57, Nb-94, enabling robustness testing. Figure \ref{f9} compares HPGe-Compton Net and Genie2000 performance for SV1$\sim$4, with vertical dotted lines marking full energy peak positions and accuracies annotated upper-right: (a) SV1: Genie2000 slightly outperformed (attributable to Poisson variance); (b) SV2$\sim$4: HPGe-Compton Net achieved superior accuracy, notably resolving SV2’s extreme low-count full energy peak undetectable by curve fitting. HPGe-Compton Net’s superior performance for real LLW samples with short measurement durations implies its high potential for industrial applications. Additional various LLW samples will be employed for further performance evaluation after the HPGe-Compton Net is upgraded by enriching the database, specifically, by including more target radionuclides and reducing its high dependence on the clean Compton continua. HPGe-Compton Net’s significantly better accuracy performance on real LLW samples with insufficient measurement duration suggests its potential industrial applicability.

\section{Conclusions}

A new Physics-Guided CNN method has been developed for gamma spectral data analysis to extract the entire response information, including the Compton continuum for each radionuclide. The HPGe-Compton Net showed five times faster measurements than the traditional methods that analyze only peak regions of the response while maintaining comparable accuracy, which was validated through performance evaluations for both synthetic datasets and real gamma spectra collected using LLW samples.

The unique methods implemented to HPGe-Compton Net include the Channel-Prompt method and the controlled-uncontaminated-Compton-edge-region database building for guiding CNN to the targeted peak location and adaptive weighting of Compton region features. Current limitations focus on the limited number of applicable radionuclide types and the requirement for a clean Compton edge region. The methodology’s success in balancing measurement speed and precision positions HPGe-Compton Net as a potential transformative tool for LLW management in the near future.

%
% Each of the commands below will create an unnumbered section with the appropriate heading.
% Remove any sections that are not relevant for your article.
% All sections except suppdata will be removed if the [anonymous] option is used.
% See iopjournal-guidelines.pdf for more information.
%

\ack{The authors would like to thank McMaster University Health Physics, especially Dave Niven, for providing equipment and radioactive sources, and McMaster Nuclear Reactor, especially Nancy Spoelstra, for helping apply the neutron activation.}

\funding{The authors would like to thank Mitacs and Laurentis Energy Partners for financially supporting this project.}
% This section is a list of funder names and grant numbers

\roles{Y X: Writing—original draft, Writing—review \& editing, Validation, Methodology, Data curation and Analysis, Modal training, Conceptualization. Y W: Writing—review \& editing, Data curation and Analysis. S B: Writing—review \& editing, Supervision, Methodology, Funding acquisition, Conceptualization.}
% List author names and the contributions made to the article, using terms from the NISO Contributor Roles Taxonomy (CRediT) https://credit.niso.org

\data{The data that support the findings of this study are openly available at the following URL/DOI: 
https://ieee-dataport.org/documents/hpge-compton-net-training-dataset}
% For more information on IOP Publishing's research data policy see: https://publishingsupport.iopscience.iop.org/questions/research-data/

%\suppdata{Sample text inserted for demonstration.}

%\section*{References}

\end{document}